\documentclass[twocolumn,aps]{revtex4}
\input{epsfig.sty}
\newcommand{\beq}{\begin{eqnarray}}
\newcommand{\eeq}{\end{eqnarray}}
\begin{document}
\title{EWPT Nucleation With MSSM and Electromagnetic Field Creation}
\author{Ernest M. Henley\\
Department of Physics, University of Washington, Seattle, WA 98195\\
Mikkel B. Johnson \\
Los Alamos National Laboratory, Los Alamos, NM 87545 \\
Leonard S. Kisslinger\\
Department of Physics, Carnegie Mellon University, Pittsburgh, PA 15213\\ }

\begin{abstract}
   Using EW-MSSM field theory, so the EWPT is first order, we derive the 
equations of motion for the gauge fields.  With an isospin ansatz
we derive e.o.m. for the electrically charged W fields uncoupled from all 
other fields. These and the lepton currents serve as the current for the 
Maxwell-like e.o.m. for the electromagnetic field. The electromagnetic 
field arising during EWPT bubble nucleation without leptons is found.  
We then calculate the electron current contribution, which is seen to be
quite large.  This provides the basis for determinining the magnetic field 
created by EWPT bubble collisions, which could seed galactic and 
extra-galactic magnetic fields.

\end{abstract}
\maketitle
\noindent
PACS Indices:12.38.Lg,12.38.Mh,98.80.Cq,98.80Hw
\vspace{3mm}
\noindent


\section{Introduction}

   There have been many attempts to find the origin of the
large-scale galactic and extra-galactic magnetic fields that have been 
observed (see Ref\cite{gr} for a review).  One possible seeding 
of these magnetic fields is via early universe phase transitions: the
electroweak phase transition (EWPT) and the Quantum Chrodynamics chiral
phase transition (QCDPT).
Research on the QCDPT has explored not only seeding of the galactic and
extra-galactic structure, but also possible effects in Cosmic Microwave
Background Radiation (CMBR) correlations\cite{co,soj,kl,bfs,dfk,fz,kv95,ae98,
cst00,son99}; and large scale magnetic structure which could produce 
observable CMBR polarization correlations\cite{lsk1} arising from bubble 
collisions during the QCDPT\cite{lsk2}.

  Since the electromagnetic (em) field along with the $W^{\pm}$ and $Z$ fields 
are the gauge fields of the Standard EW model, the EWPT is particularly 
interesting for exploring possible cosmological magnetic seeds . It has been
shown that in the Standard EW model there is no first order EWPT\cite{klrs}, 
no explanation of baryogenesis, or any interesting cosmological magnetic
structures created during the crossover transition. However, there has
been a great deal of activity in the supersymmetric extension of the 
standard model\cite{r90}, and with a MSSM having a Stop with a mass 
similar to the Higgs there can be a first order phase transition and 
consistency with baryogenesis\cite{laine,cline,losada}. 

   Moreover, with a first order phase transitions bubbles of the new phase 
nucleate and collide, which is essential for our present research on magnetic 
field creation.  A number of authers have used EW-MSSM models with the 
Standard EW Model plus a right-handed Stop to explore the EWPT. See, e.g., 
Refs. \cite{laine,bodeker}. There have been other models for CP violation and 
baryogenesis, such as two-Higgs models (see Refs.\cite{laine,cline,losada} 
for references and discussion) and leptoquarks (see, e.g., Ref.\cite{herczeg} 
for a discussion and references.)

 The goal of our present work on the creation of electromagnetic fields via 
nucleation is both to find the nature of the fields and to derive the
boundary conditions for our study of magnetic field generation with EWPT 
bubble collision, which is in progress\cite{sjkhhb08,sj09}. For this reason
we introduce the complete Lagrangian that is needed to study bubble
collisions. In our present calculations we could use a two-Higgs model, as 
the precise nature of the extension of the Standard EW Model is not needed 
for creation of electromagnetic fields during nucleation, as we shall 
show; while in our studies of magnetic field generation with EWPT bubble 
collision we  use the MSSM and treat all the equations of motion, derived 
below, with a right-handed Stop.

In the most detailed previous research on the study of magnetic fields 
arising from EWPT transitions\cite{kv95,ae98,cst00}  a model was used in 
which the equations of motion (e.o.m.) for the em field involved the chargeless
Higgs, and the EWPT bubbles were empty of the em field until the bubbles 
collided and overlapped. In the present work we derive the e.o.m.
directly from the EW Lagrangian, using a MSSM including the right-handed Stop 
field for consistency with a first order phase transition. The e.o.m. which 
we obtain for the em field is Maxwell-like, with the current given by the
charged $W^{\pm}$ fields, which is physically reasonable.  

  In general, the e.o.m. are complicated coupled partial differential 
equations, in which solutions for the Higgs and Stop fields must be found to
obtain the current for the em field equation. In the present paper, however, 
we introduce an I-spin formulation which allows us to uncouple the e.o.m. for 
the $W^{\pm}$ fields, and carry out a study of spherically symmetric EW bubble 
nucleation. For pure nucleation of a bubble we first solve the $W^{\pm}$ 
e.o.m. numerically without leptons, and obtain solutions for the em field 
using a fit to these numerical solutions for the $W^{\pm}$ fields. Next we 
include the electron current and solve the new equations for the 
electromagnetic field.

  The solutions have an instanton-like form near the bubble wall,
as expected. These are the first solutions for the electromagnetic field
from EWPT nucleation that have been obtained from the EW Lagrangian.
Due to spherical symmetry during nucleation only electric fields are
produced.  These give the initial fields for bubble collisions.
For the derivation of EW cosmological magnetic seeds, the
more complicated collision problem must also be solved\cite{sjkhhb08}

  In Section II we present the EW-MSSM model, including fermions, and
derive the e.o.m. for the gauge fields and the Higgs and Stop fields.
In Section III the I-Spin ansatz is used to separate the gauge fields,
and we present the solutions for the em field without fermions. In
Section IV we include electrons, the mmost important fermion for
em field creation, and find the em field. Our conclusions are given
in Section V

\section{MSSM EW equations of motion with right-handed stop}

In this section we derive the equations of motion for the
standard Weinberg-Salam EW model extended to a MSSM model
with the addition of a Stop field, the partner to the top quark.
all other partners of the standard model fields are integrated out.

\subsection{MSSM EW Lagrangian with Fermions}

\beq
\label{L}
  {\cal L}^{MSSM} & = & {\cal L}^{1} + {\cal L}^{2}  + {\cal L}^{3} 
+ {\cal L}^{fermion}  \\
     {\cal L}^{1} & = & -\frac{1}{4}W^i_{\mu\nu}W^{i\mu\nu}
  -\frac{1}{4} B_{\mu\nu}B^{\mu\nu}  \nonumber \\
 {\cal L}^{2} & = & |(i\partial_{\mu} -\frac{g}{2} \tau \cdot W_\mu
 - \frac{g'}{2}B_\mu)\Phi|^2  -V(\Phi) \nonumber \\
 {\cal L}^{3} &=& |(i\partial_{\mu} -\frac{g_s}{2} \lambda^a C^a_\mu)\Phi_s|^2
    -V_{hs}(\Phi_s,\Phi) \nonumber \\
    {\cal L}^{fermion}& = & {\rm \;standard \;Lagrangian \;for \;fermions}
 \nonumber \; ,
\eeq
where the pure $C^a_\mu$ term is omitted in ${\cal L}^{1}$ and  
\beq
\label{wmunu}
  W^i_{\mu\nu} & = & \partial_\mu W^i_\nu - \partial_\nu W^i_\mu
 - g \epsilon_{ijk} W^j_\mu W^k_\nu\\ \nonumber
 B_{\mu\nu} & = & \partial_\mu B_\nu -  \partial_\nu B_\mu \, ,
\eeq
where the $W^i$, with i = (1,2), are the $W^+,W^-$ fields, $C^a_\mu$
is an SU(3) gauge field, ($\Phi$, $\Phi_s$) are the (Higgs, right-handed 
Stop fields), $(\tau^i,\lambda^a)$ are the (SU(2),SU(3) generators, and
the electromagnetic and Z fields are defined as 
\beq
\label{AZ}
   A^{em}_\mu &=& \frac{1}{\sqrt{g^2 +g^{'2}}}(g'W^3_\mu +g B_\mu) \nonumber \\
   Z_\mu &=& \frac{1}{\sqrt{g^2 +g^{'2}}}(g W^3_\mu -g' B_\mu) \; .
\eeq 

   The standard fermion Lagrangian, ${\cal L}^{fermion}$ has the form
\beq
\label{fermion}
      {\cal L}^{fermion} & = & \sum_j[\bar{\psi}^j_L\gamma^\mu(i\partial_{\mu} 
-\frac{g}{2} \tau \cdot W_\mu  - g'\frac{Y}{2}B_\mu) \psi^j_L  \nonumber \\
      && +\bar{\psi}^j_R\gamma^\mu(i\partial_{\mu} -g'\frac{Y}{2}B_\mu) 
\psi^j_R] \nonumber \\
      && + \sum_c f^i \bar{\psi}^c_L \Phi \psi^c_R \; ,
\eeq
with j representing all fermions, while the sum over c is just for charged
fermions. The first term in ${\cal L}^{fermion}$ gives the interaction of
fermions with the gauge fields, while second term corresponds to the Higgs
EWPT creation of fermion masses. We neglect these masses, as well as fermion
masses created by the Stop, in the present work.
 
\subsection{MSSM EW Equations of Motion Without Fermions}

The effective Higgs and Stop potentials are taken as
\beq
\label{V}
    V(\Phi) & = & -\mu^2 |\Phi|^2 + \lambda |\Phi|^4 \nonumber \\
    V_{hs}(\Phi,\Phi_s) & = & -\mu^2_s |\Phi_s|^2 + \lambda_s |\Phi_s|^4 \\
              &&  +\lambda_{hs} |\Phi|^2 |\Phi_s|^2 \, .\nonumber
\eeq
The various parameters are discussed in many publications\cite{laine}. In 
particular we
need $g=e/sin\theta_W = 0.646$, $g'=g \; tan\theta_W =0.343$, and 
$G=gg'/\sqrt{g^2 +g^{'2}}=0.303$.

In the picture we are using, the Higgs and Stop fields will play a dynamic
role in the EW bubble nucleation and collisions, and we shall 
need the space-time structure of these fields rather than only
the vacuum expectation value for a particular vacuum state for
the complete solutions of the e.o.m.
Our form for the Higgs field, $\Phi$, is 
\beq
\label{phi}
         \Phi(x) & = & \left( \begin{array}{clcr} 0 \\
                                \phi(x)
         \end{array} \right) \; .
\eeq
and 
\beq
\label{tau}
  \tau \cdot W_\mu \Phi & = & 
    \left( \begin{array}{clcr} (W^1_\mu-iW^2_\mu) \\
                            - W^3_\mu
         \end{array} \right)  \phi(x) \, .
\eeq

In the present exploratory paper treating bubble nucleation
we center on the possible generation of an electromagnetic field,
and the solution of all of the e.o.m. is avoided. 
Therefore, specific forms and solutions for the Higgs and Stop fields
do not enter the equations needed for the present
work. For this reason we do not choose a specific form for the right-handed
Stop field, $\Phi_s$, and for convenience write the additional MSSM gauge
field as
\beq
\label{cmu}
              C_\mu &=& \frac{\lambda^a}{2} C^a_\mu \; .
\eeq
We also use the definitions
\beq
\label{phis}
       \phi(x) &\equiv& \rho(x)e^{i\Theta(x)} \nonumber \\
       |\phi(x)|^2 &=& \rho(x)^2 \\ 
        |\Phi_s(x)|^2 & \equiv & \rho_s(x)^2 \nonumber \; .
\eeq         
Although we do not need specific forms for  $C_\mu$ or $\Phi_s$,
we assume that a Stop condensate is formed for consistency with a first order
EWPPT, as in Ref \cite{bodeker}

With these definitions $ {\cal L}^{2}$ is (j = (1,2,3)) 
\beq
\label{L2}
 {\cal L}^2 & = &  \partial_\mu \phi^*\partial^\mu \phi +
(i(\partial_\mu \phi^*)\phi - i\phi^* \partial_\mu \phi)
( -g W^3_\mu \nonumber \\
          && +g' B_\mu)/2 + \phi^* \phi((\frac{g}{2})^2 (W^j)^2 +
(\frac{g'}{2})^2 B^2 \nonumber \\
 &&-\frac{gg'}{2} W^3 \cdot B)-V(\phi) \; .
\eeq 

The equations of motion for the charged gauge fields, $W^{i}$, for i=(1,2)  
are obtained by minimizing the action
\beq
\label{action}
 \delta \int d^4 x [{\cal L}^{1}+{\cal L}^{2} +{\cal L}^{3}] & = & 0 \, ,
\eeq
i.e., we do not include ${\cal L}^{fermion}$.
The equations of motion that we obtain from the variations in 
$W^{i}$, for i=(1,2) are
\beq
\label{eom1}
  && \partial^2 W^i_\nu-\partial^\mu \partial_\nu  W^i_\mu
 -g \epsilon^{ijk} {\cal W}^{jk}_\nu
 +\frac{g^2}{2}\rho(x)^2 W^i_\nu \nonumber \\
     && \;\;\; = 0 \; ,
\eeq
with 
\beq
\label{w's}
     {\cal W}^{jk}_\nu &\equiv&  \partial^\mu( W^j_\mu) W^k_\nu  
     + W^j_\mu \partial^\mu  W^k_\nu + W^{j\mu} W^k_{\mu\nu}\;, \nonumber \\
     W^k_{\mu\nu}&=& \partial^\mu W^k_\nu-\partial^\nu W^k_\mu-g\epsilon^{klm}
  W^l_\mu W^m_\nu
\eeq
The e.o.m. for $A^{em},Z$, including the eletron em current, $J^e_\nu$, are
\beq
\label{eom2}
 &&\partial^2 A^{em}_\nu-\partial_\mu \partial_\nu  A^{em}_\mu
 -\frac{gg'}{\sqrt{g^2 + g^{'2}}} \epsilon^{3jk} {\cal W}^{jk}_\nu \nonumber \\
 && -J^e_\nu\;\;\;\;\; = 0
\eeq
\beq
\label{eom3}
 &&\partial^2 Z_\nu-\partial_\mu \partial_\nu Z_\mu -
\frac{\rho^2\partial_\nu \Theta}{\sqrt{g^2 + g^{'2}}} \nonumber \\
&&  -\frac{g^2}{\sqrt{g^2 + g^{'2}}} \epsilon^{3jk} {\cal W}^{jk}_\nu 
  = 0\;.
\eeq
Note that we have dropped all fermion currents except the electron current.

 The e.o.m. for the Higgs field are
\beq
\label{eom4}
 &&\frac{1}{\rho(x)} \partial^2 \rho(x) -\mu^2+ 2\lambda \rho(x)^2 +
\lambda_{hs} \rho_s(x)^2  -H \cdot H \nonumber \\
&&-\partial_\mu \Theta \partial^\mu \Theta 
+\frac{\sqrt{g^2 + g^{'2}}}{2} Z^\mu \partial_\mu \Theta = 0 \; ,
\eeq
with
\beq
   H \cdot H &\equiv&  
  (\frac{g}{2})^2 W^i \cdot W^i
+(\frac{g'}{2})^2 B \cdot B -\frac{gg'}{2} W^3 \cdot B \nonumber\; ,
\eeq
and
\beq
\label{eom5}
 \partial_\mu (\rho(x)^2 \partial^\mu \Theta -\frac{\sqrt{g^2 + g^{'2}}}{2}
  \rho(x)^2 Z^\mu ) &=& 0 \; .
\eeq
The e.o.m. for the right-handed Stop is 
\beq
\label{stop}
  &&-\partial^2 \Phi_s +ig_S(\partial^\mu(C_\mu\Phi_s)+(C_\mu\partial^\mu\Phi_s))
 +(g_s^2 C_\mu^\dag C_\mu \nonumber \\
  && +\mu_s^2+2\lambda_s\rho_s^2 +\lambda_{hs}\rho^2)
\Phi_s = 0 \; ,
\eeq
We do not give the e.o.m. for the $C_\mu$ gauge field, which is not needed in 
the present work.

These are exact equations of motion in our MSSM model with a right-handed stop,
without fermions, which (as we shall see in the next Section) have only small 
effects In our research on EWPT collisions, where we calculate the magnetic 
field produced by the EWPT, we need the equations for the Higgs and Stop
fields, but in our present work on nucleation we only calculate the
gauge fields, with the inlusion of the eletron current given in the following
section.
\section{I-Spin Ansatz and Electromagnetic Field Creation from $W^{\pm}$}

   One of the most important features of the equations of motion derived
directly from the EW Lagrangian is that the source current  of the 
electromagnetic field is given by the charged gauge $W^{\pm}$ fields,
as seen from the Maxwell-like Eq.(\ref{eom2}). This is expected physically,
and is in sharp contrast with the Kibble-Vilenkin\cite{kv95}, 
Ahonen-Enqvist\cite{ae98}, Copeland-Saffin-T$\ddot{o}$rnkvist\cite{cst00} 
picture. This suggests that we use an SU(2), isospin ansatz for the gauge 
fields. 

\subsection{I-Spin Ansatz}

  In the present paper we assume that 
\beq
\label{nuci1}
  W^j_\nu &\simeq& i\tau^j W_\nu(x) \simeq i\tau^j x_\nu W(x)\;\; {\rm j=1,2,3}
 \nonumber \\
  A^{em}_\nu &\simeq& i\tau^3 A_\nu(x)\simeq i\tau^3 x_\nu A(x) \nonumber \\
  Z_\nu &\simeq& i\tau^3 Z(x)_\nu \simeq i\tau^3 x_\nu Z(x) \; , 
\eeq
with the I-spin operators defined as $\epsilon^{mjk} \tau^j\tau^k = i\tau^m$.
We shall see that this enables us to derive the straight-forward
equations of motion for the electromagnetic field, which can be solved
to a good approximation for symmetric nucleation of EW bubbles. In this 
section we derive the e.o.m. for spherically symmetric bubble nucleation, 
so that $W(x)=W(r,t)$ and $A(x)=A(r,t)$,
with $x^\mu x_\mu = t^2-r^2$. First note that
\beq
\label{Wjk}
     \epsilon^{ijk} {\cal W}^{jk}_\nu &=& i\tau^i \times F[W_\nu,
\partial_\nu W]\; ,
\eeq
with $F$ a function of $W_\nu$ and $\partial_\nu W$ to be 
determined, so that the e.o.m. for $W_\nu$, Eq.(\ref{eom1}), becomes
\beq
\label{eom6}
  && \partial^\mu \partial_\mu W_\nu-\partial_\nu \partial_\mu W^\mu
-gx_\nu [5 W^2+  \nonumber \\
 && 3 W(t\partial_t+ r\partial_r)W +g s^2 W^3  -\beta s^2 \partial_r W]
\nonumber \\ 
 && -\frac{g^2}{2} \rho^2 W_\nu = 0 \; ,
\eeq
with $W_\nu = x_\nu W(r,t),\; s^2 = t^2-r^2,\; r=\sqrt{\sum_{j=1}^{3} x^j x^j}
= \sqrt{-\sum_{j=1}^{3} x^j x_j},\; \partial_j r= x^j/r$, and $\beta  = (+,-)$ 
for $\nu = (t,j)$. Subtracting 
the e.o.m. for $W_t \times x_j$ from the e.o.m. for $W_j \times t$ we find 
\beq
\label{eomW}
  && (\partial_t^2 + \partial_r^2)W +\frac{t^2+r^2}{rt}\partial_t
 \partial_r W +(\frac{3}{r}\partial_r + \frac{3}{t}\partial_t)W \nonumber \\
  &&+gW (t^2-r^2)(\frac{1}{t}\partial_t-\frac{1}{r}\partial_r)W = 0 \\
\label{eomA}
   && (\partial_t^2 + \partial_r^2)A +\frac{t^2+r^2}{rt}\partial_t
 \partial_r A +(\frac{3}{r}\partial_r + \frac{3}{t}\partial_t)A \nonumber \\
  && +GW (t^2-r^2)( \frac{1}{t}\partial_t-\frac{1}{r}\partial_r)W = 0 \; .
\eeq

The most significant aspect of the I-spin formulation is that we obtain
e.o.m. for $W(r,t)$ and for $A(r,t)$ without contributions from the Higgs 
or Stop fields.

As pointed out at the beginning of this section, the current for the 
electromagnetic field arises entirely form the electrically charged 
fields/particles, $W^{\pm}$, as seen also in Eq.(\ref{eomA}). Moreover,
the current within our I-spin formulation is determined by a nonlinear
partial equation for $W(r,t)$, without direct coupling to the Higgs or
Stop fields. This will enable us to derive the electromagnetic fields
from EWPT bubble collisions as in Refs.\cite{kv95,ae98,cst00}. In
the present paper, however, we derive the electromagnetic fields
produced in the EWPT via bubble nucleation before collisions, which
has not been considered previously. For collisions a direction in space
is singled out, so that the form $W(r,t),A(r,t)$ cannot be used. This
is a topic for future work. Also, fermions contribute to the electric current,
and fermion fields will be included in future work.
The solution for $A^{em}$ produced during
nucleation with the assumptions of the present section are found in the 
following section.

\subsection{I-Spin Ansatz and Electromagnetic Field Creation During Nucleation
Without Fermions}

   In the present work we make use of the gauge fields gauge conditions to
reduce the partial differential equations, Eqs(\ref{eomW},\ref{eomA}), 
to ordinary differential equations. The philosophy is to derive the $W^{\pm}$
and $A^{em}$ fields as a function of r at a fixed time. Since from the
general structure of the equations we expect at time t that the bubble
wall will be at $r\; =\; r_w\;\simeq \;t$, we are mainly interested in the 
nature of
the fields near r = $r_w$. As we shall see, since the solutions are 
modified instanton-like in nature, the most significant region for magnetic
field creation for both nucleation, and for collisions, will be at the
bubble walls. 

  We use the Coulomb gauge, which is consistent with spherical spatial 
symmetry, giving the equations for $W(r,t)$ and $A(r,t)$:
\beq
\label{coulgauge}
   \sum_{j=1}^{3} \partial_j W^j &=& \sum_{j=1}^{3} \partial_j A^j = 0
 \;\; {\rm or} \nonumber \\
    r\partial_r W(r,t) + 3 W(r,t) &=& 0\;,  
\eeq
with solutions for each value of r
\beq
\label{gsoln}
    W(r,t) &=& \frac{W_r(t)}{r^3}\;\; A(r,t) = \frac{A_r(t)}{r^3}\; .
\eeq

   From Eq.(\ref{eomW}) and the gauge condition (\ref{coulgauge}), 
one obtains differential equations for $W^t(r,t)$ and $W^j(r,t)$. By
combining them one finds that the Higgs and Stop fields are disconnected,
and obtains an e.o.m. for the functions $W_r(t)$ and $A_r(t)$
\beq
\label{eomWt}
  W_r''(t) - \frac{3t}{r^2} W_r'(t) +\frac{3}{r^2} W_r(t) +g\frac{t^2-r^2}{r^3}
W_r(t) \\
   (\frac{1}{t}W_r'(t)-\frac{3}{r^2}W_r(t)) &=& 0 \nonumber\\
\label{eomAt}
 A_r''(t) - \frac{3t}{r^2} A_r'(t) +\frac{3}{r^2} A_r(t) +G\frac{t^2-r^2}{r^3}
W_r(t) \\
 (\frac{1}{t}W_r'(t)-\frac{3}{r^2}W_r(t)) & =& 0 \nonumber
\eeq

We proceed by 1) finding initial conditions and numerical solutions to 
Eq.(\ref{eomWt}) for W(t) for a series of r-values, 2) fitting a function 
to these  values, and 3) finding the function
\beq
\label{H(t)}
      H(t) &=& \frac{t^2-r^2}{r^3}W(t)(\frac{1}{t}W'(t)-\frac{3}{r^2}W(t)),
\eeq
which is used in Eq.(\ref{eomAt}) to obtain an approximate solution for
A(t) and thereby A(r,t), using Eq.(\ref{gsoln}).

In Figure 1 $W_r(t)$ is given for various values of r,
and the time for the creation of the bubble wall is clearly seen.
In Figure 2 similar results are shown for $A_r(t)$.
Note that near the bubble wall $A^{em}$ becomes infinite and has an 
instanton-like behavior
\beq
\label{instanton}
     A(r,t)|_{r \simeq r_{wall}} &\simeq& \frac{A_W}{((r^2 - t^2) +\zeta^2)^2} \; .
\eeq 
Away from the surface 
A(t) becomes smaller than this instanton-like solution.   

\begin{figure}[ht]
\epsfig{file=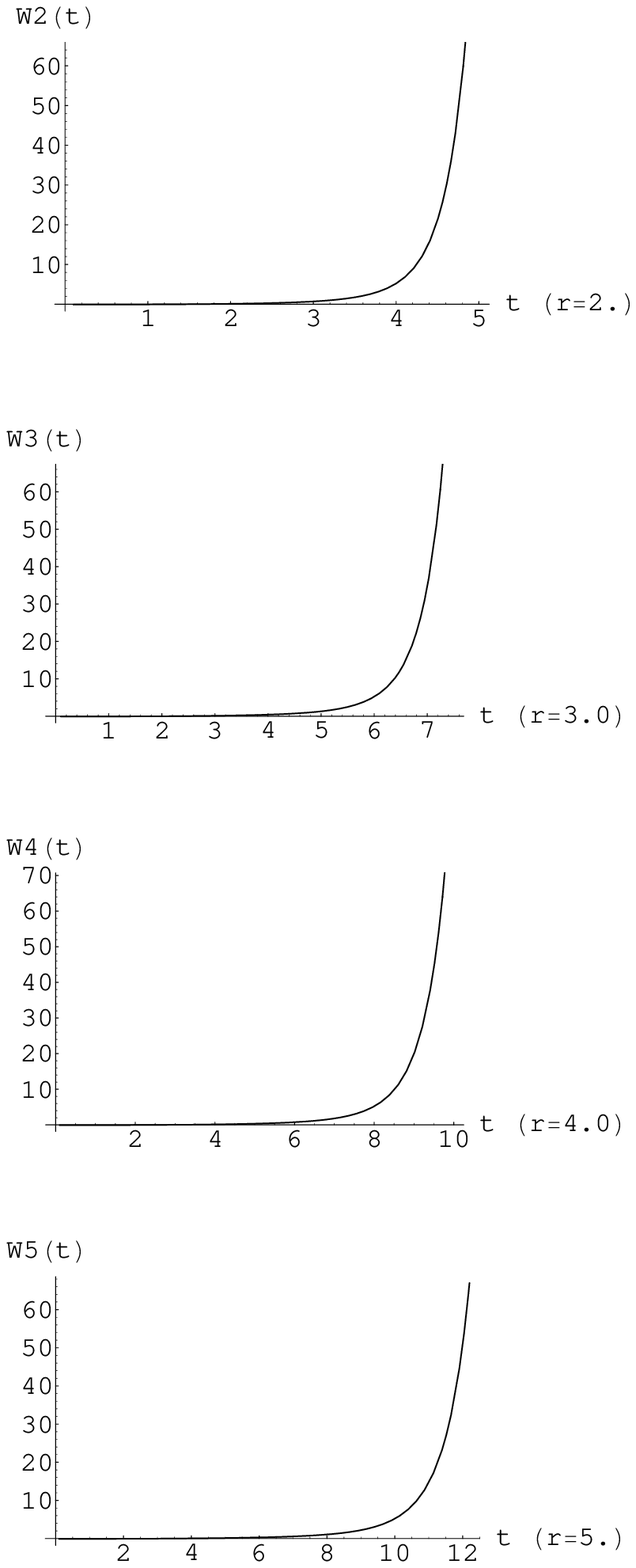,height=15cm,width=6.cm}
\caption{The function $W_r(t)$ for various values of r}
{\label{Fig.1}}
\end{figure}
\vspace{5cm}

\begin{figure}[ht]
\epsfig{file=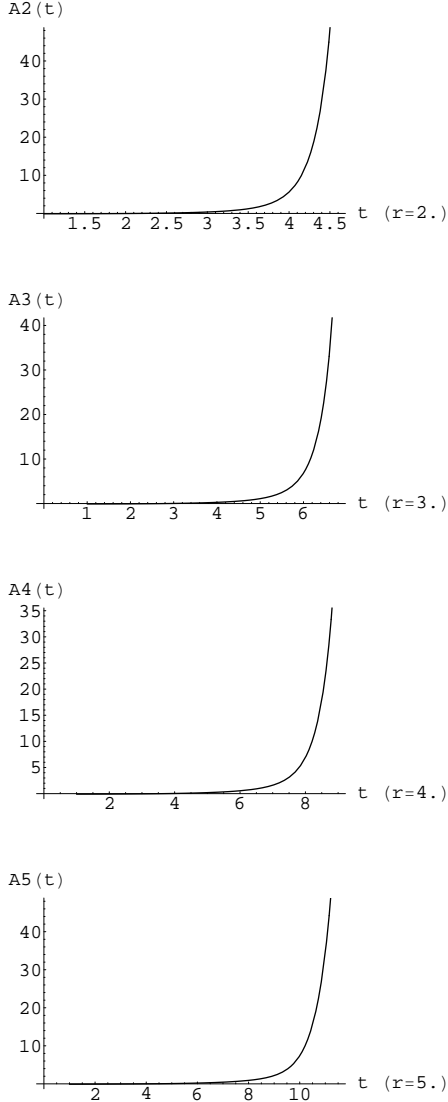,height=15cm,width=6.cm}
\caption{The function $A_r(t)$  for various values of r}
{\label{Fig.2}}
\end{figure}

   From Figure 2 one observes that the time at which one reaches the radius 
of the wall bubble is given approximately by
\beq
\label{26}
             t & \simeq & 2r \; ,
\eeq
from which we obtain the nucleation velocity of the bubble wall.
\beq
\label{27}
             v^{wall} & \simeq & \frac{c}{2} \; .
\eeq
This is an important result in our work on magnetic field generation
and evolution, as well as our recent study of gravity wave production
by magnetic fields created in the EWPT\cite{kks09}

In our present work $A^{em}_{\nu} \sim x_{\nu}A(r,t)$, so electric
fields but no magnetic fields are created. This work can provide the initial
conditions for EWPT collisions in which magnetic fields are created.  

\section{Electromagnetic Field Creation During Nucleation With I-Spin 
Ansatz and Fermions}

   Let us now return to Eq(\ref{eom2}), the e.o.m. for the em field with
the electron current as well as the charged gauge fields
\beq
 &&\partial^2 A^{em}_\nu-\partial_\mu \partial_\nu  A^{em}_\mu
 -\frac{gg'}{\sqrt{g^2 + g^{'2}}} \epsilon^{3jk} {\cal W}^{jk}_\nu \nonumber \\
 && -J^e_\nu\;\;\;\;\; = 0 \nonumber \; ,
\eeq
with the electron current at finite temperature\cite{kt}

\beq
\label{jelectron}
     J^{lep}_\nu &=&G n_e \bar{u}_e \gamma_\nu u_e \equiv j \bar{u}_e 
\gamma_\nu u_e  \nonumber \\
              n_e &=& \frac{3}{4\pi^2} \zeta(3)T^3 \; ,
\eeq
with $ \zeta(3)\simeq 1.2$, G=$gg'/\sqrt{g^2+g'^2}$=0.303, and $u_e$ a
Dirac spinor.

  From Eq(\ref{eom2}) we obtain the e.o.m. for $A_r(t)$ (see previous
section)
\beq
\label{eom7}
      A''-\frac{3 t}{r^2}A'+\frac{3}{r^2}A +G H -3 r^2 j &=& 0\; ,
\eeq
with $j=0.028 M_{Higgs}^3$. We have taken $T=M_{Higgs}$ at the time of
the EWPT. 

    The solutions for $A_r(t)$ for fixed r are shown in Figs. 3 and 4.
As one can see, the electron current plays an important role
in electromagnetic field production during nucleation of EWPT bubbles.
Note that the current of the charged W is approximately -$jW_r(t)$.
The solutions for $jW_r(t)$ for fixed r, in comparison to the 
electron current, are shown in Fig 5. The function $jWr(t)$ should be 
compared to the electron current = -$0.084 r^2$.
Although the electron current is much smaller than the current from
the $W^+,W^-$ fields near the bubble surface, the electron
current is larger than the $W^+,W^-$ currents in the interior of the
bubble.. 

  Since the electron current is larger than the charged gauge currents
inside the bubble, as shown in Fig. 5, we conclude that the leptonic
currents must be included for the correct derivation of the magnetic fields 
during EWPT bubble collisions.

\clearpage

\begin{figure}
\begin{center}
\epsfig{file=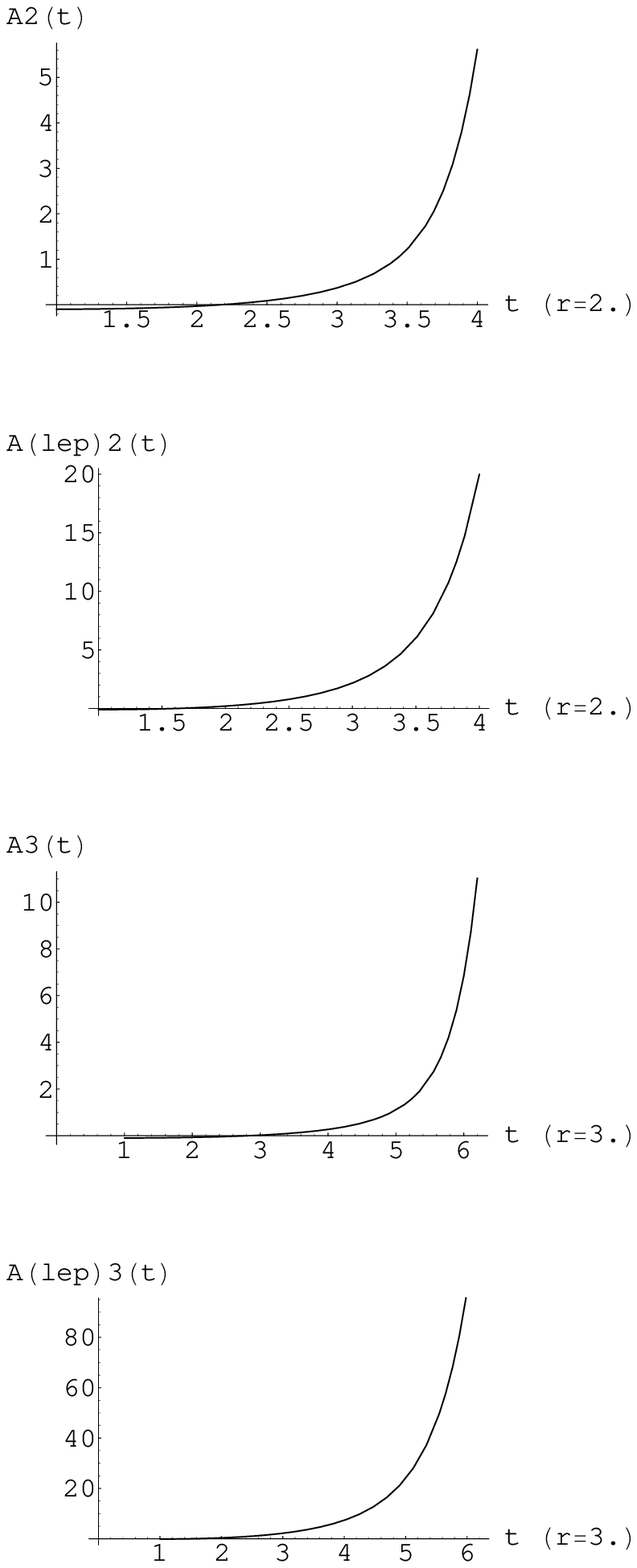,height=15cm,width=6cm}
\caption{A(t) without and A(lep)(t) with lepton currents for r=2.0 and 3.0}
{\label{Fig.3}}
\end{center}
\end{figure}

\begin{figure}
\begin{center}
\epsfig{file=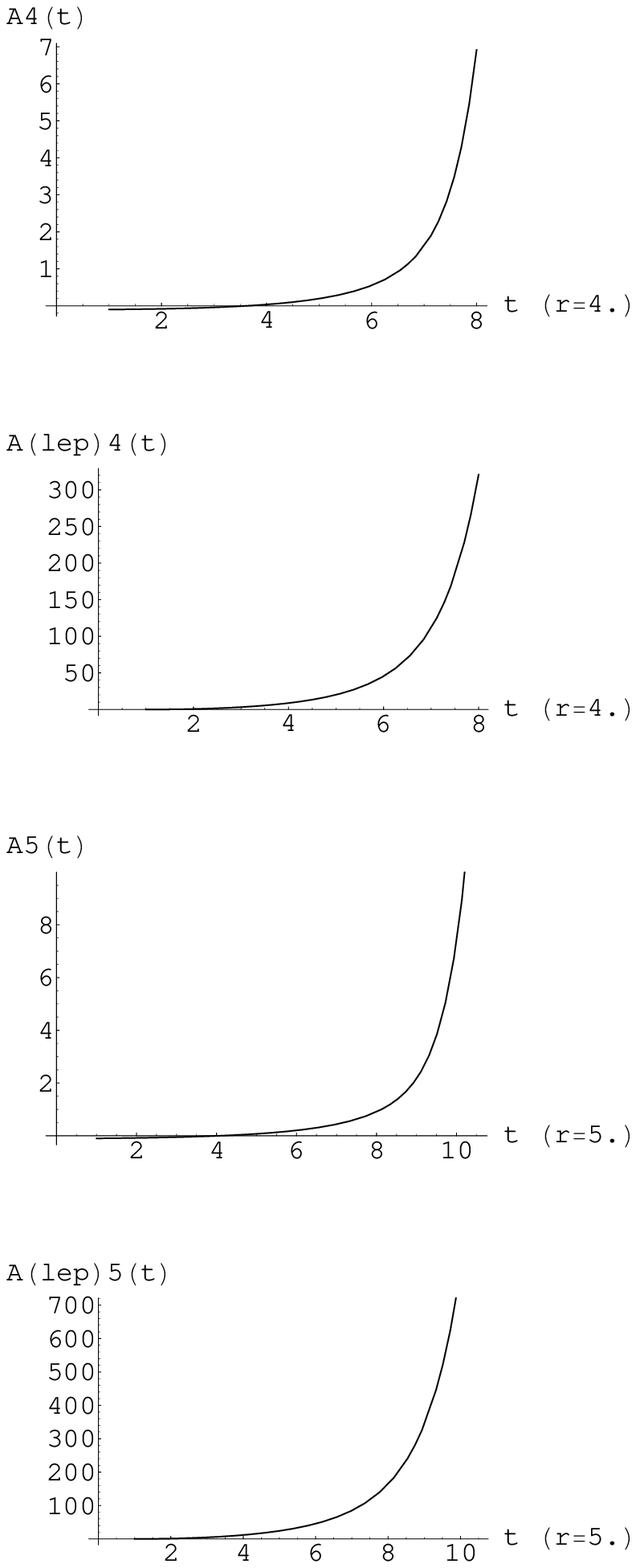,height=15cm,width=6cm}
\caption{A(t) without and A(lep)(t) with lepton currents for r=4.0 and 5.0}
{\label{Fig.4}}
\end{center}
\end{figure}

\begin{figure}
\begin{center}
\epsfig{file=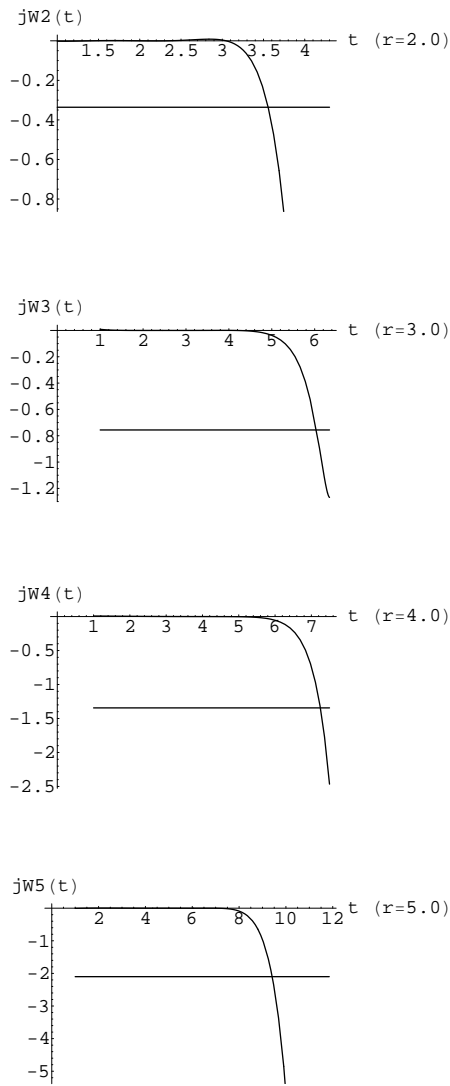,height=15cm,width=6cm}
\caption{jWr=W current for r=2.0, 3.0, 4.0, and 5.0 The horizontal lines are 
the electron current = -0.084 r$^2$}
{\label{Fig.5}}
\end{center}
\end{figure}

\section{Conclusions}

   We have formulated the coupled equations of motion for the electroweak
MSSM with a Lagrangian that adds the right-handed Stop field terms to the 
Standard Model.

In this model a first order EWPT can occur with satisfactory baryogenesis.
By using an I-spin ansatz for spherically symmetric bubble nucleation we
were able to derive a Maxwell-like equation of motion for the electromagnetic
field with the current given by the electrically charged gauge fields, 
$W^{\pm}$, as well as the charged fermion currents. Moreover,
by treating the $W_t$ and $W_j$ components separately  we were able to 
decouple the equations for the $W^{\pm}$ fields from the other gauge fields,
and also the Higgs and Stop fields, and obtain the current for the 
electromagnetic field.

In the present paper we derived solutions for the electromagnetic
field caused by EWPT bubble nucleation, using a Coulomb gauge condition
to obtain ordinary differential equations, from which we found instanton-like
solutions for the electromagnetic field in the region of the bubble wall.
We calculated the electromagnetic field produced in the EWPT bubble
nucleation with electron currents as well as the $W^{\pm}$ fields, and found
that although the electron current did not affect the em field near the
bubble wall, they play an important role within the bubble, which could be
quite important for bubble collisions. 

Although our present work is a very limited physical problem, it explores 
new physics which can arise from nucleation before collisions starting 
from a MSSM electroweak Lagrangian with leptons. This will be investigated 
in a continuation of our current research\cite{sjkhhb08,sj09} on magnetic
field creation in EWPT bubble collisions. In the future we shall investigate
whether the resulting magnetic fields could serve as seeds for gallactic
and extra-gallactic large-scale magnetic fields.

\vspace{3mm}

\Large{{\bf Acknowledgements}}\\
\normalsize
This work was supported in part by the NSF grant PHY-00070888, in part 
by the DOE contracts W-7405-ENG-36 and DE-FG02-97ER41014. 
The authors thank Prof. W.Y. Pauchy Hwang and Dr. S. Walawalkar for helpful 
discussions, and Los Alamos National Laboratory for hospitality.
\vspace{2cm}



\end{document}